\documentclass[conference]{IEEEtran}
\IEEEoverridecommandlockouts
\usepackage{cite}
\usepackage{amsmath,amssymb,amsfonts}
\usepackage{algorithmic}
\usepackage{graphicx}
\usepackage{textcomp}
\usepackage{algorithm}
\usepackage{multirow}
\usepackage{diagbox}
\usepackage{caption}
\usepackage{subfigure}

\def\BibTeX{{\rm B\kern-.05em{\sc i\kern-.025em b}\kern-.08em
    T\kern-.1667em\lower.7ex\hbox{E}\kern-.125emX}}
\begin{document}

\title{Fast Community Detection based on Graph Autoencoder Reconstruction\\}

\author{\IEEEauthorblockN{Chenyang Qiu}
\IEEEauthorblockA{cyqiu@bupt.edu.cn}
\and
\IEEEauthorblockN{Zhaoci Huang}
\IEEEauthorblockA{zc.huang@bupt.edu.cn}
\and
\IEEEauthorblockN{Wenzhe Xu}
\IEEEauthorblockA{wenzhexu@outlook.com}
\and
\IEEEauthorblockN{Huijia Li}
\IEEEauthorblockA{hjli@bupt.edu.cn}
\and
\\
\IEEEauthorblockA{~~~~~~~~~~~~~~~~~~~~~~~~~~~~~~~~~~~~~~~\textit{School of Science} \\
~~~~~~~~~~~~~~~~~~~~~~~~~~~~~~~~~~~~~~~~~~~~~~~~~~\textit{Beijing University of Posts and Telecommunications}\\
~~~~~~~~~~~~~~~~~~~~~~~~~~~~~~~~~~~~~~~~~Beijing, China\ \\
}
}
\maketitle

\begin{abstract}
With the rapid development of big data, how to efficiently and accurately discover tight community structures in large-scale networks for knowledge discovery has attracted more and more attention. In this paper, a community detection framework based on \textbf{G}raph \textbf{A}uto\textbf{E}ncoder \textbf{R}econstruction (noted as GAER) is proposed for the first time. GAER is a highly scalable framework which does not require any prior information. We decompose the graph autoencoder-based one-step encoding into the two-stage encoding framework to adapt to the real-world big data system by reducing complexity from the original $O(N^2)$ to $O(N)$. At the same time, based on the advantages of GAER support module plug-and-play configuration and incremental community detection, we further propose a peer awareness based module for real-time large graphs, which can realize the new nodes community detection at a faster speed, and accelerate model inference with the 6.15$\times$-14.03$\times$ speed. Finally, we apply the GAER on multiple real-world datasets, including some large-scale networks. The experimental result verified that GAER has achieved the superior performance on almost all networks.
\end{abstract}
\
\newline
\indent
\begin{IEEEkeywords}
Big data mining, Community detection, Graph autoencoder, Data inference, Large graphs
\end{IEEEkeywords}

\section{Introduction}
As one of the most important physical tools to portray the real world, network data today is gradually developing in the direction of large-scale, complexity, and modularity. For example, the network of social platforms, the protein interaction network in genetic engineering, and the transportation network, power grid, etc. There are not only complex interactions between nodes, the network will also form different communities due to this interaction and node attributes. Topologically, community can be understood that internal nodes are relatively tightly connected and externally connected relatively sparse. Identifying this local structure is essential for understanding complex systems and knowledge discovery~\cite{181}.

The above task is community detection. There have been a lot of researches on community detection, including optimization based methods~\cite{Li1}~\cite{Jin04}, label propagation methods~\cite{LP}. And another is the reconstruction based method. 

In 2016, Yang proposed a nonlinear reconstruction method based on the autoencoder (denoted as DNR) for the first time~\cite{Yang}. After 2018, on the basis of this work, a variety of autoencoder reconstruction methods that integrate different network features have been proposed~\cite{Caob}~\cite{Bha}~\cite{Caoa}. However, these methods often require additional operations and have limited ability to capture network features, so that the $Q$ value of this DNR method is very poor when targeting a network with an unknown community structure. We empirically demonstrate this phenomenon in VI.

On the other hand, the graph neural network based community detection was first proposed in 2019~\cite{cdgnn}~\cite{Shc}. However, there are currently a little works, and most of them are based on non-backtracking theories, Markov Random Field, etc.~\cite{Chen}~\cite{Jin}, which are supervised methods. However, the prior information (such as labels) of the network community for big data systems is sometimes rare, which also causes a great challenge to above semi-supervised methods. So it is very emergent to propose an unsupervised method for this field.

Based on the above questions, we proposed a more efficient and accurate community detection framework named as GAER based on graph autoencoder reconstruction. Our innovative contributions are mainly as follows:
\begin{itemize}
\item GAER is a novel and brand new graph neural network based community detection model designed for big data system. We redesigned the input features, nonlinear activation function and loss function, as well as the corresponding downstream tasks. GAER has achieved significantly performance improvement than DNR and graph autoencoder (GAE) in the experimental stage, which shows that GAER is not a trivial extension based on GAE.
\item A joint optimization framework based on modularity and network structure by graph autoencoder reconstruction is proposed for the first time in community detection. GAER significantly achieves the best results on almost small or large-scale networks whether it has a known community structure or not. And this performance improvement benefits from nonlinear modularity reconstruction and neighborhood Laplacian smooth sampling of our model, which can partially mitigate the extreme degeneracy problem and resolution limit caused by single modularity maximization.
\item In order to achieve fast community detection, a two-stage encoding framework is proposed by further decomposing the one-step encoding based on GAE, which significantly reduces the complexity of GAER from $O(N^2)$ to $O(N)$. And this two-stage operation is easy to deploy in parallel or distributed. Then benefits from the non-parametric decoder design, GAER can reduce half of the parameters learning in some configurations and perform integrated back propagation learning instead of training each encoder separately like DNR. These advantages also improve algorithm efficiency and scalability.
\item The GAER framework implements incremental community detection and the most advanced module configuration. So we propose a feature-Aligned Peer Awareness Module (APAM) for faster community detection of new nodes in real-time large-scale networks. Experiments show that while ensuring the accuracy of community detection, APAM will speed up node inference by $\times$6.15-$\times$14.03, which has a strong practical significance for big data systems built on networks nowadays.

\end{itemize}
\section{Related work}


\subsection{Reconstruction and Modularity maximum}
This model was first introduced by Newman to maximize the modularity index $Q$ of the network~\cite{Newman}, which is defined by the following:
\begin{equation}\label{maxq}
Q=\frac{1}{2 M} \sum_{i, j}\left[\left(a_{i j}-\frac{k_{i} k_{j}}{2 M}\right) \delta\left(\sigma_{i}, \sigma_{j}\right)\right],
\end{equation}
where $M$ is the total number of edges of the network, $a_{i j}$ is the adjacency matrix element, a value of 1 or 0 indicates whether there is a connected edge or not, $k_{i}$ is the degree of node $i$, and $\delta$ is the association membership function of node $i$, when $i$ and $j$ belong to the same community, $\delta=1$, otherwise $\delta=0$.

Then, we simplify the Eq.(\ref{maxq}) by defining the module degree matrix $\mathbf{B}$ and introducing the node community membership vectors. Define the module degree matrix $\mathbf{B}=\left[b_{i j}\right]$ as
\begin{equation}
b_{i j}=a_{i j}-\frac{k_{i} k_{j}}{2 m}.
\end{equation}
In this way, each node has a modular degree relationship with all the other nodes, whether they have connected edges or not.

Next, we can get a matrix $\mathbf{H}=\left[h_{i j}\right] \in \mathbb{R}^{N \times K}$ which each row $h_i$ is the community membership vector, and $K$ is the dimension of the node community membership vector. So the Eq.~(\ref{maxq}) can be reduced as the following:
\begin{equation}\label{Q}
Q=\frac{1}{2 m} \operatorname{Tr}\left(\mathbf{H}^{\mathrm{T}} \mathbf{B H}\right).
\end{equation}

There are many different optimization ways to solve the maximization of Eq.(\ref{Q}), but as a NP hard problem, here we introduce $\mathbf{H}^{\mathrm{T}} \mathbf{H}$ as the constant $N$ condition to relax the problem, so we obtain the following modularity optimization problem after the relaxation:
\begin{equation}
\begin{aligned}
\max Q=\max \left\{\operatorname{Tr}\left(\mathbf{H}^{T} \mathbf{B H}\right)\right\}\\
\text { s.t. } \operatorname{Tr}\left(\mathbf{H}^{\mathrm{T}} \mathbf{H}\right)=N.
\end{aligned}
\end{equation}
Based on the Rayleigh entropy, we know that the solution $\mathbf{H}$ of the modularity degree maximization problem under relaxation conditions is the $k$ largest eigenvectors of the modularity degree matrix $\mathbf{B}$. Then, according to Eckart and Young's matrix reconstruction theorem, the equivalence of the modular degree maximization and the matrix reconstruction can be obtained.

\section{A universal framework for community detection: GAER}
\subsection{Graph autoencoder}
The graph autoencoder was proposed by Kipf in 2016~\cite{Kipf} and the encoder was specifically defined as follow:
\begin{equation}
\mathbf{Z}=\operatorname{GNN}(\mathbf{X}, \tilde{\mathbf{A}}),
\end{equation}
and the $\tilde{A}$ is as follow:
\begin{equation}
\tilde{\mathbf{A}}=\mathbf{D}^{-\frac{1}{2}}\mathbf{(A+I)} \mathbf{D}^{-\frac{1}{2}}.
\end{equation}

The authors recommended encoding using a two-layer GCN in the following:
\begin{equation}
\mathbf{Z}=\mathrm{GCN}(\mathbf{X}, \widetilde{\mathbf{A}})=\widetilde{\mathbf{A}} \operatorname{ReLU}\left(\tilde{\mathbf{A}} \mathbf{X} \mathbf{W}_{0}\right) \mathbf{W}_{1}.
\end{equation}

And the decoder is as follow, then a cross entropy loss can be used to minimize the reconstruction loss between $\mathbf A$ and $\hat{\mathbf{A}}$:
\begin{equation}
\hat{\mathbf{A}}=\sigma\left(\mathbf{Z Z}^{\mathrm{T}}\right).
\end{equation}

\begin{figure*}
  \centering
  \includegraphics[scale=0.23]{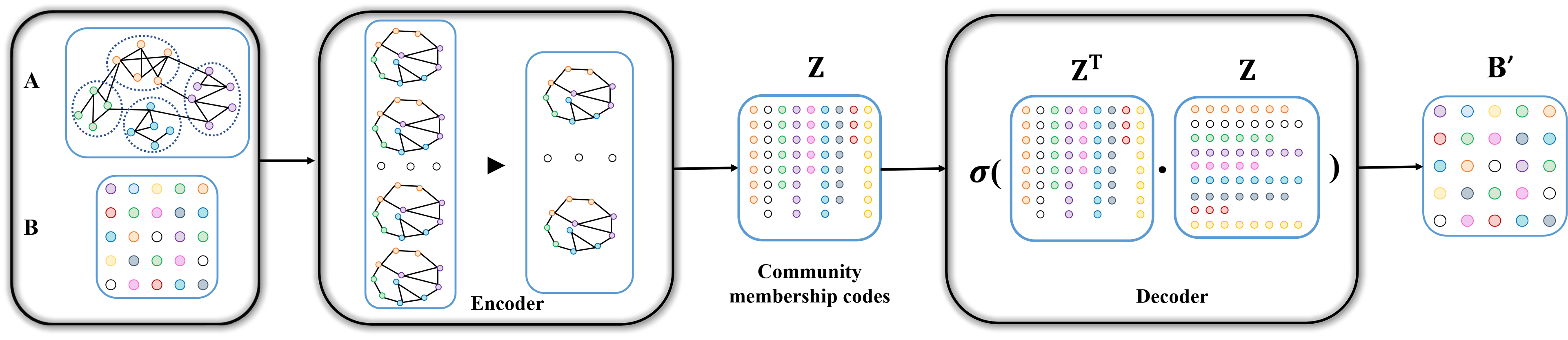}\\
  \caption{\small The architecture of the community detection model GAER. We use the adjacency matrix $\mathbf A$ and modularity information $\mathbf B$ on the left (or concatenate the original features $\mathbf X$ of the nodes) to encode node community membership $\mathbf Z$. The modularity matrix is reconstructed by the decoder on the right to maximize the modularity.}
\end{figure*}
\subsection{GAER-GCN}
Inspired by graph autoencoder, we propose a community detection model based on modularity reconstruction: GAER-GCN, which jointly uses the modularity and the network structure.
\subsubsection{Model}
\
\newline
\indent GAER is also available to use node features, so for the input features of the first layer $\mathbf{B}^{0}$, we have:
\begin{equation}
\mathbf{B}^{0}=\operatorname{CONCAT}(\mathbf{B},\mathbf{X}),
\end{equation}
where $\mathbf X$ is the original feature of the node, and CONCAT means the concatenating operation. If the network has no node feature, then $\mathbf{B}^0=\mathbf{B}$.

Then, we use a graph convolutional network (GCN) as an encoder to jointly use the modularity information and structural information of the network via Eq.(\ref{GAER-GCN}) to obtain the best node community membership vector. We also redesign the reconstruction index and loss function in the following. This is why the model is called GAER-GCN.

\begin{equation}\label{GAER-GCN}
\mathbf{b}^{l}=\operatorname{GCN}\left(\mathbf{B}^{l-1}, \mathbf{A}\right)=\operatorname{tanh}\left(\widetilde{\mathbf{A}} \mathbf{B}^{l-1} \mathbf{W}_{l}\right).
\end{equation}
${b}^{l}$ is the low-rank community membership matrix encoded by $i$-th layer. And tanh is the activation function. It is particularly important to note that we use tanh for the GAER activation function. We must point out the necessity of replacing ReLu with tanh, because there are a large number of negative elements in the modularity matrix $\mathbf B$. If ReLu is applied after the fully connected layer in GAER, negative elements will be set to 0 by ReLu, which will cause the gradient of this element to always be 0, and the weight cannot be updated. This will  greatly reduces the accuracy of GAER.

And the reconstruction task is
\begin{equation}
\hat{\mathbf{B}}=\sigma(\mathbf{b}^{L}\left(\mathbf{b}^{L}\right)^{T}).
\end{equation}

\subsubsection{Optimization}
\
\newline
\indent Note the parameter space is $\delta=\left\{\mathbf{W}_{1}, \mathbf{W}_{2}, \ldots, \mathbf{W}_{\mathrm{L}}\right\}$, the optimization task is as follow:
\begin{equation}
\begin{aligned}
\hat{\delta} &=\underset{\delta}{\arg \min } \mathcal{L}_{\mathrm{Re}}(\mathbf{B}, \hat{\mathbf{B}})=\underset{\delta}{\arg \min } \sum_{i=1}^{N} \mathcal{L}_{\mathrm{Re}}\left(b_{i}, \hat{b}_{i}\right) \\
&=\underset{\delta}{\arg \min } \sum_{i=1}^{N} \mathcal{L}_{\mathrm{Re}}\left(b_{i}, \mathrm{GCN}\left(b_{i}, a_{i}\right)\right),
\end{aligned}
\end{equation}
where $\mathcal{L}_{\mathrm{Re}}$ is a distance function, with either Euclidean distance or cross-entropy distance.

Here, we specially design the corresponding F-norm loss for the fast community detection task, which uses the Euclidean distance between two matrices.
\begin{equation}
\begin{aligned}
\hat{\delta} &=\underset{\delta}{\arg \min } \mathcal{L}(\mathbf{B}, \hat{\mathbf{B}})=\underset{\delta}{\arg \min }\left\|\mathbf{b}^{L}\left(\mathbf{b}^{L}\right)^{T}-\mathbf{B}\right\|_{\mathrm{F}}^{2} \\
&=\underset{\delta}{\arg \min } \sum_{i=1}^{N} \sum_{j=1}^{N}\left(\hat{b}_{i j}-b_{i j}\right)^{2}.
\end{aligned}
\end{equation}
And this loss function performs well in subsequent experiments, and can be further added a $l_2$ regularization to constrain the reconstruction. We leave $l_2$ regularization and another cross-entropy loss function based on sigmoid distance for future work.

\subsubsection{The properties}
\
\newline
\indent GAER-GCN has realized the optimal community detection via above model. However GAER-GCN obtains a low-rank membership matrix through a one-step encoding by matrix multiplication. We find that the complexity of this operation derived from the original graph autoencoder is too high, and it is $O(N^2)$, which is very detrimental to the extension on big data systems. Therefore, we propose a two-stage GAER framework which algorithm can be reduced to $O(N)$ to improve the scalability of our model on the big data systems nowadays and accelerate the community detection on large graphs.

\subsection{The framework of GAER}
Analyzing the one-step encoding of GAER-GCN, it can be found that GAER-GCN obtains neighborhood information by multiplying the adjacency matrix and the feature matrix. However, the adjacency matrix is a sparse matrix, and a large number of 0 elements multiplied by features will only lead to high complexity of the algorithm.

Therefore, we decompose one-step encoding into two-stage encoding: Neighborhood Sharing and Membership Encoding, through a limited number of neighbor sampling $k$; we will obtain condensed and high-quality neighborhood information.

The Neighborhood Sharing stage is as follow:
\begin{equation}\label{NS}
b_{n(v)}^{l} = N S\left(\left\{b_{u}^{l-1}, \forall u \in n(v)\right\}\right).
\end{equation}
$NS$ is a Neighborhood Sharing operator. We use the MEAN as $NS$, and MEAN can add and average the node neighbor information. The other operators are recommended as LSTM, Pool~\cite{SAGE} and GAT~\cite{GAT} ect.

And the Membership Encoding stage is as follow:
\begin{equation}\label{ME}
b_{v}^{l} = \sigma\left(\operatorname{CONCAT}\left(b_{v}^{l-1}, b_{n(v)}^{l}\right) \mathbf{\cdot W}_{l}\right).
\end{equation}

The pseudo code of the two-stage GAER framework is showed in Algorithm~\ref{alg1}. 
\begin{algorithm}
	\renewcommand{\algorithmicrequire}{\textbf{Input:}}
	\renewcommand{\algorithmicensure}{\textbf{Output:}}
	\caption{Framework of the two-stage GAER}
	\label{alg1}
	\begin{algorithmic}[1]
        \REQUIRE  Graph $G(V, E)$; node features $\left\{\mathrm{x}_{v}, \forall v \in V\right\} ;$ layer depth $L$; weight matrices $\mathbf{W}_{l}, \forall l \in\{1,2, \ldots, L\}$; non-linearity $\sigma$; neighborhood node set: $n(v)$.
        \ENSURE  Community membership vectors ${z}_{v}$ for all $\forall v \in V$
		\FOR  {$l$ =1,2,...,$L$}
		\FOR {$\forall v \in V$}
		\STATE \textbf{Neighborhood Sharing}: Sample the neighborhood codes of node $v$, and obtain neighbor information $b_{n(v)}^{l}$ by Eq.(\ref{NS})
		\STATE \textbf{Membership Encoding}: Merge of $v$ and its neighborhood, and obtain low-rank membership encoding $b_{v}^{l}$ by Eq.(\ref{ME})
		\ENDFOR
		\ENDFOR
		\STATE   $\mathbf{z}_{v} \leftarrow \mathbf {b}_{v}^{l}, \forall v \in V$
	\end{algorithmic}
\end{algorithm}

As can be seen in the next stage. The two-stage GAER (noted as GAER) also provides incremental community detection and can configure more advanced GNN modules on it.

\subsection{Can we faster? Toward real-time large graphs: GAER-APAM}
Through a two-stage encoding framework, we reduced the GAER complexity from $O(N^2)$ to $O(N)$, a complexity in a linear relationship with the network nodes. However, as a complex giant system, the real-world network has not only a large number of nodes, but the addition of new nodes in real time. GAER can infer the new nodes with $O\left(2k^{L} N\right)$ complexity. However when each layer of GAER increases, the inferred complexity will increase exponentially, which greatly increases the cost of time and is detrimental to real-time incremental community detection. Therefore, we propose a module APAM to achieve faster and better real-time community detection.
\subsubsection{Node feature alignment}
\
\newline
\indent Since the scale of the network that realizes incremental prediction is very large, suppose the number of nodes that grow in a period of time is m, and the original number of nodes in the network is n, then $n>>m$, that is, the addition of m will not significantly affect the original large graph structure. At the same time, it is time-consuming and unnecessary to add a new node to link all nodes in the network and obtain its modularity information. Therefore, we propose a new node modularity information alignment strategy, aiming to replace the modularity information of the new node by selecting the most similar neighborhood node modularity information of the new node. And then concatenate the inherent features and modularity information of the newly added node as the inference input features.
\begin{algorithm}
	\renewcommand{\algorithmicrequire}{\textbf{Input:}}
	\renewcommand{\algorithmicensure}{\textbf{Output:}}
	\caption{Feature alignment}
	\label{alg2}
	\begin{algorithmic}[1]
        \REQUIRE  New node $v_i$; the original feature of $v_i$: $\mathbf x_i$; the $k$ neighbors sampling set of $v_i$: $N_{v_i}$; the $k$ neighbors sampling set of $N_{v_i}$: $N_{j}^{v_i}~\{j={1,2,...,k\}}$; the number of common neighbors: $t$; the index of most similar node: $curr$
        \ENSURE  the GEAR feature of $v_i$: $\mathbf {X}_i$
        \STATE Initialize $t = 0$; $curr = 0$
		\FOR  {$j$ =1,2,...,$k$}
        \IF {$(N_{j}^{v_i} \cap N_{v_i}) > t$}
        \STATE $curr = j$
        \STATE $t=~(N_{j}^{v_i} \cap N_{v_i})$
        \ENDIF
		\ENDFOR
		\STATE   $\mathbf {b}_{i} \leftarrow (\mathbf b_{curr}^{0}$ CONCAT $\mathbf {x}_{i}$)
	\end{algorithmic}
\end{algorithm}

\subsubsection{Aligned Peer Aware Module(APAM)}
\
\newline
\indent Each extended layer of GAER will consume much memory and time for neighbor sampling, thus a peer aware module PAM was proposed that yields better results at lower layer numbers. Specifically, before the Neighborhood Sharing, an attention aggregation of neighbor nodes is done which can be seen as a type of self attention aggregation in the node neighborhood set. Such operations indirectly capture as much deep neighborhood information as possible in the case of one layer. Let the $v$ have K neighbors $\left\{\mathbf{b}_{i}^{l}\right\}_{i=1}^{K}$, the neighbor nodes after passing through the PAM are expressed as~\cite{Yan}:
\begin{equation}\label{PAM}
\mathbf{b}_{i}^{l^{*}}=\sum_{j} \alpha_{i j} \mathbf{b}.
\end{equation}

We can obtain the self-attention coefficient $\alpha_{i j}$ through the softmax assignment of the standardized dot product in the neighbor set:
\begin{equation}\label{self-attention}
\quad \alpha_{i j}=\frac{\exp \left(a_{i j}\right)}{\sum_{k} \exp \left(a_{i k}\right)} , \quad a_{i j}=\frac{\left\langle\mathbf{b}_{i}, \mathbf{b}_{j}\right\rangle}{\sqrt{d}}.
\end{equation}

Combined with our proposed node feature alignment strategy for new nodes in large-scale networks, we further propose a high-speed inference module APAM. By first aligning the new node features, then operating peer aware learning. Specific algorithm pseudo code is shown in Algorithm\ref{alg3}.

As for node inference, we firstly obtain the network community structure through a high-layer GAER offline training, and then fine-tune and train a low-layer inference model on the premise of obtaining low-layer weights. This part of the time is a low-order infinitesimal in complexity compared to the high-layer model, so it will cost less of time. But the more accurate community structure can be obtained by the high-layer model, which will make the further incremental prediction more accurate.
\begin{algorithm}
	\renewcommand{\algorithmicrequire}{\textbf{Input:}}
	\renewcommand{\algorithmicensure}{\textbf{Output:}}
	\caption{GAER-APAM (one layer)}
	\label{alg3}
	\begin{algorithmic}[1]
        \REQUIRE  New node $v_i$; the GAER feature of $v_i$: $\mathbf b_i$; the neighbors of $v_i$: $N_{v_i}$; the GAER feature of neighbors: $\mathbf b_v ,v \in N_{v_i}$
        \ENSURE  the membership of $v_i$ 
        \STATE Obtain the feature of $v_i$ by Algorithm~\ref{alg2}
		\FOR {$\forall v \in N_{v_i}$}
        \STATE Update the $\mathbf b_{v}^{0}$ by Eq.~(\ref{PAM}) and Eq.~(\ref{self-attention})
		\ENDFOR
		\STATE Obtain the $v_i$'s neighbor information $\mathbf b_{n(v)}^{1}$ by Eq.~(\ref{NS})
        \STATE Obtain $v_i$'s membership encoding $\mathbf z_{v_i}^{1}$ by Eq.~(\ref{ME})
        \STATE Obtain $v_i$'s community membership by incremental K-means clustering of $\mathbf z_{v_i}^{1}$
	\end{algorithmic}
\end{algorithm}
\subsection{Computational complexity analysis}

\begin{table*}[t]
\scriptsize
\centering
\caption{Multi-layer model complexity analysis.}
\begin{center}
\begin{tabular}{|c|c|c|c|}
\hline
\textbf{Algorithms} & \textbf{\textit{Encoding Stage}}& \textbf{\textit{Decoding Stage}}& \textbf{\textit{Total Complexity}} \\
\hline
GAER(one-stage)& $O(p N^{2}+(2 N-1) d p )+ o(N^2)$& $O(d p N)$& $O (p N^{2}+3 d p N-d p )+ o(N^2) \sim \boldsymbol{O}\left(\boldsymbol{N}^{2}\right)$ \\
GAER(two-stage)& $O((k+2d) p N- d p)+o(N)$ & $O(d p N)$ & $O((3 d+k) p N-d p )+o(N) \sim \boldsymbol{O}(\boldsymbol{N})$\\
DNR& $O(2 d p N)+o(N)$ & $O(2 d p N)$ & $O(4 d p N)+o(N)\sim \boldsymbol{O}(\boldsymbol{N})$\\
GAER(inference)& $(k^L-1) N + k^{(L-1)} (2N-1) d + o(k^LN)$&-& $(k^L-1) N + k^{(L-1)} (2N-1) d + o(k^LN)\sim \boldsymbol{O}(\boldsymbol{2k^LN})$\\
APAM& $O(kdN+k(k-1)d^2+k^2)$& - &$O(kdN+k(k-1)d^2+k^2) \sim \boldsymbol{O}(\boldsymbol{dkN})$ \\
\hline
\multicolumn{4}{l}{Where $k$ is the number of samples in the neighborhood, $d$ is the dimension of the low-dimensional embedding, $p$ is the number of samples in each minibatch.}\\
\multicolumn{4}{l}{And $k, d, p \ll N$, can be regarded as a constant. Here only shows the APAM module complexity. '-' means there is no such operation complexity.}\\

\end{tabular}
\label{complexity}
\end{center}
\end{table*}

 For algorithms applied to real-world data, complexity analysis is crucial. We analyze the algorithm complexity of the multi-layer model. Specially, limited to the inherent over-smoothing and model degradation problems of GNN~\cite{TO}, the multi-layer in GAER is generally 2-5 layers. How to redesign GAER to fit the deep architecture of GNN is a interesting issue, we leave this for future work.

 Lines 1-3 of the Table~\ref{complexity} show the forward propagation complexity of the GAER and DNR mini-batch training models. In order to simplify the complexity formulas, we set input feature dimension in the first layer as $N$, and the output feature dimension as $d$ ($d<<N$), and after the first layer the input and output feature dimension are all $d$. It can be seen from Table~\ref{complexity} that the main component of the forward complexity of the training model is concentrated in the first layer. Because $N$-dimensional embedding is required, the complexity of the remaining layers is briefly replaced by higher-order infinitesimals; while the complexity of the single node inference stage of GAER is mainly concentrated In the one-level aggregation and embedding involving $L$-hop neighbors, the rest is replaced by higher-order infinitesimals. Finally we show that the APAM module is a linear operation, then the GAER with the APAM module is used to infer the complexity of $O(dkN+2k ^LN)$. Intuitively, the complexity of the inference model with APAM added is higher than that of the original inference model. However, in high-layer models, GAER-APAM will be a very competitive choice, because the complexity of the inference model will grow exponentially by layer. APAM can achieve as much accuracy as possible with less time complexity. This is very practical on real-world big data systems.

\begin{figure}[h]
  \centering
  \includegraphics[scale=0.5]{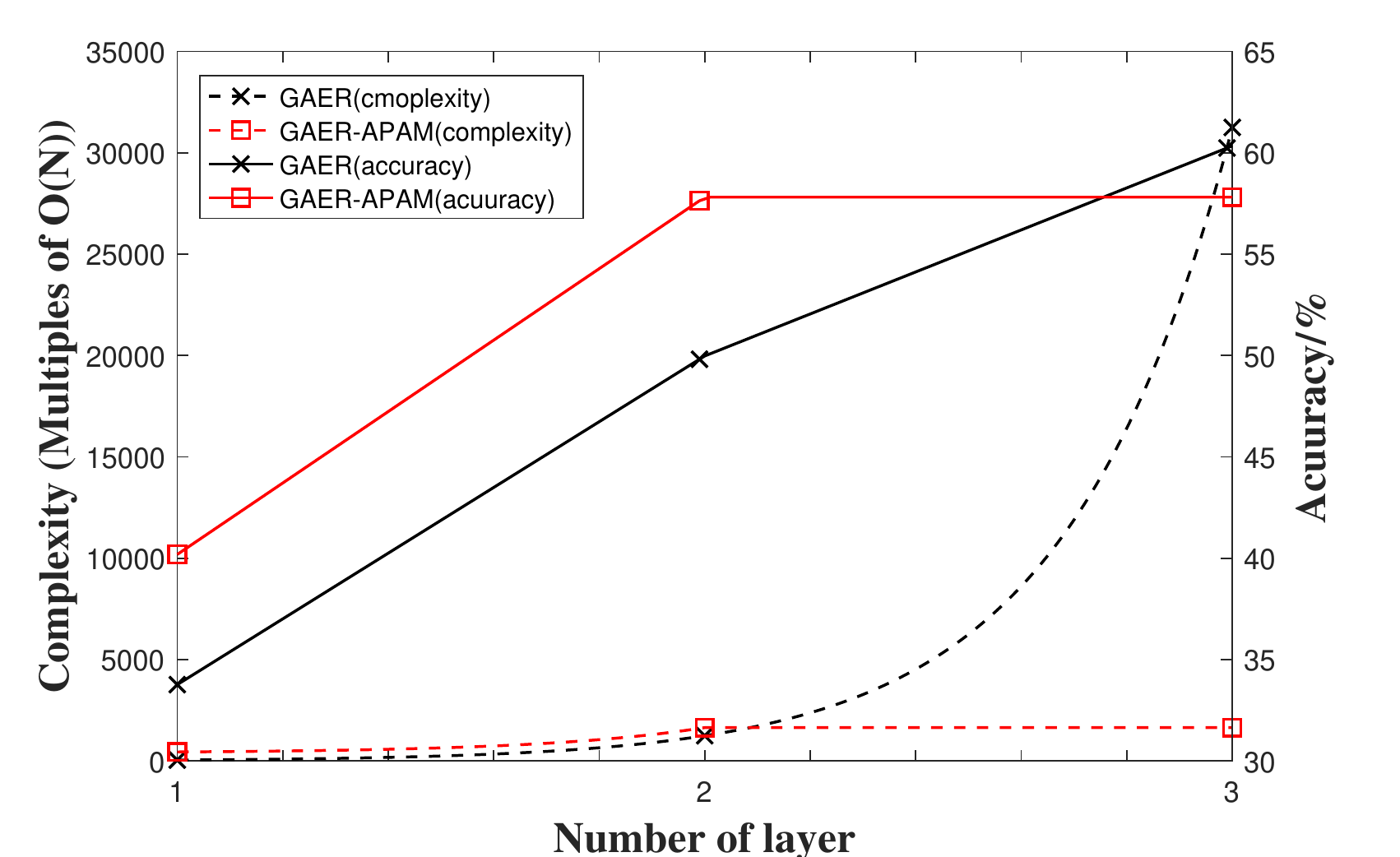}\\
  \caption{The comparison of complexity and inference accuracy in the different layers of GAER and GAER-APAM.}\label{complexity_fig}
\end{figure}

 Fig.\ref{complexity_fig} shows this clearly by contrast the accuracy and complexity of two models. The red line in the figure represents GAER with APAM module, which we continue to use a two-layer configuration in the three-layer structure. As can be seen from the Fig.\ref{complexity_fig}, the complexity of the first two layers of GAER-APAM is slightly more linear than that of GAER, but the inference accuracy is significantly improved. For the third layer, the complexity of GAER has increased exponentially, but only a small increase in accuracy has been obtained.

\section{Experiment}
\subsection{Experiment on real classic networks}

\begin{table}[h]
\center
\caption{\small Eight real-world networks and their related properties, where $K$ is the preset number of communities, $N$ is the number of nodes, and $M$ is the number of edges}
\label{Clsssic}
\begin{tabular}{cccccc}
\hline\noalign{\smallskip}
Symbol & Dataset & K & N & M & C \\
\noalign{\smallskip}\hline\noalign{\smallskip}
$G_1$ & Karate & 2 & 34 & 78 & K \\
$G_2$ & Dolphins & 2 & 62 & 159 & K \\
$G_3$ & Friendship & 6 & 68 & 220 & K \\
$G_4$ & Football & 12 & 115 & 613 & K \\
$G_5$ & Polblogs & 2 & 1490 & 16718 & K \\
$G_6$ & Cora & 7 & 2708 & 5429 & K \\
$G_7$ & Les Miserables & - & 77 & 254 & U \\
$G_8$ & Adjnoun & - & 112 & 425 & U \\
$G_9$ & Netscience & - & 1589 & 2742 & U \\
$G_{10}$ & PPI & - & 2361 & 7182 & U \\
$G_{11}$ & Power Grid & - & 4941 & 6594 & U \\
$G_{12}$ & Lastfm\_asia & - & 7624 & 27806 & U \\
\noalign{\smallskip}\hline
\end{tabular}
\end{table}

We selected 12 real-world network datasets, as shown in Table~\ref{Clsssic}, and half of them have unknown community structure which separately have their respective characteristics, thus giving the algorithms different challenges to more comprehensively evaluate the effects of each algorithm. Among them, $G_6$ and $G_{10}$ are the real networks with node features, and they are large and complex structure. The other networks have no feature. 

As for comparison algorithms, we choose other 5 kinds of methods, which are representative and can be divided into two categories: one is the most advanced non-deep learning community detection method (2019-2020): they are graph embedded based GEMSEC~\cite{GEMSEC}, multi-objective evolutionary based algorithm RMOEA~\cite{RMOEA}; the other is deep learning-based nonlinear method: autoencoder reconstruction based DNR~\cite{Yang}, deep autoencoder-like nonnegative matrix factorization method DANMF~\cite{DANMF} and deep neural based method GAE~\cite{Kipf}, they are all unsupervised.

\begin{table}[h]
\center
\caption{NMI values of 7 algorithms on 6 classic networks with known community structure}
\label{Clsssic_result}
\begin{tabular}{cccccccc}
\hline\noalign{\smallskip}
 & GAER & RMOEA & GEMSEC & DANFM & DNR & GAE \\
\noalign{\smallskip}\hline\noalign{\smallskip}
$G_1$ & \textbf{1}     & \textbf{1}     & \textbf{1} & 0.833 & \textbf{1}     & 0.892 \\
$G_2$ & \textbf{0.893} & 0.703 & 0.632 & 0.661 & 0.532 & 0.597 \\
$G_3$ & \textbf{0.932} & 0.602 & 0.689 & 0.573 & 0.794 & 0.813 \\
$G_4$ & 0.893 & \textbf{0.902} & 0.734 & 0.467 & 0.544 & 0.625 \\
$G_5$ & \textbf{0.764} & 0.597 & 0.372 & 0.413  & 0.337 & 0.422 \\
$G_6$ & \textbf{0.411} & 0.334 & 0.231 & 0.357 & 0.308 & 0.408 \\
\noalign{\smallskip}\hline
\end{tabular}
\end{table}

\begin{table}[b]
\center
\caption{$Q$ values of 6 algorithms on 6 classic networks with unknown community structure}
\label{Q_result}
\begin{tabular}{cccccccc}
\hline\noalign{\smallskip}
 & GAER & RMOEA & GEMSEC & DANFM & DNR & GAE \\
\noalign{\smallskip}\hline\noalign{\smallskip}
$G_7$ & \textbf{0.5529}     & 0.4430     & 0.4760 & 0.2854 & 0.4354     & \emph{0.4872} \\
$G_8$ & \textbf{0.2873} & 0.0800 & \emph{0.1900} & 0.0410 & -0.0026 & 0.0038 \\
$G_9$ & \emph{0.8705} & \textbf{0.9062} & 0.7580 & 0.5560 & 0.3099 & 0.5348 \\
$G_{10}$ & \textbf{0.6464} & \emph{0.4213} & 0.3921 & 0.1237 & 0.2254 & 0.3328 \\
$G_{11}$ & \textbf{0.7021} & \emph{0.6827} & 0.6991 & 0.6410 & 0.1452 & 0.3638 \\
$G_{12}$ & \textbf{0.7288} & 0.3546 & 0.3272 & 0.2549 & 0.2740 & \emph{0.5595} \\
\noalign{\smallskip}\hline
\end{tabular}
\end{table}

Table.\ref{Clsssic_result} shows the NMI results on known-community-structure networks, GAER achieves the best standardized mutual information value NMI on most real networks, and the only one $G_4$ that does not reach the optimal effect is only 1\% difference. This shows the ability of GAER to correctly divide nodes with known community structures. One of them worth mentioning is $G_6$. $G_6$ is a network with node features, but most comparison algorithms can not use node feature including DNR. So most algorithms perform poorly on $G_6$, which sufficiently shows that the advantage of GEAR that take both of modularity and node feature into consideration.

Table~\ref{Q_result} shows the $Q$ value results for the 6 networks with unknown community structure, GAER achieved the best $Q$ value among the 5 networks, and the other one was slightly lower than the highest $Q$ value around 3.9\%. This shows that GAER can efficiently and accurately find the tight module structure (community) in the network. On the other hand, the semi-supervised method will completely fail in this kind of network, while GAER can find close communities and other interesting structures.

 Among them, $G_8$ and $G_{10}$ are worth mentioning. $G_8$ is a small network, But compared to $G_7$, $G_8$'s community structure is more difficult to detect. Other algorithms have $Q$ values around 0 to 0.1, while GAER reaches the highest $Q$ value, which makes GAER's Q value improvement on $G_8$ significantly higher than its improvement on $G_7$, because the community structure of $G_7$ is easier to detect; $G_{10}$ is the protein interaction network with node feature, which is very complex and has disconnected parts. This leads to poor detection results of other algorithms in the community. GAER can reach 0.6464, which greatly exceeds the effects of other algorithms. This fully shows the superiority of joint optimization with modularity and the network features. At the same time, we horizontally compare DNR and GAE, as part of the inspiration for the GAER algorithm. First of all, the $Q$ value of DNR is very poor in various unknown-community networks, and each result during the experiment is very unstable and oscillates in a large range. This phenomenon illustrates DNR uses autoencoders to mechanically reconstruct $\mathbf B$, and the network structure and node features cannot be used. On the other hand, GAE performs slightly better in various networks, which suggests that in community detection tasks, the use of network structure still has great utility, but the effect of GAE is still far worse than GAER. As for the $G_8$ network, the $Q$ value of GAE is almost 0, the community result cannot be detected. Excluding $G_8$, GAER has an improvement of 13.49\%-94.23\% over GAE, which fully proves the effectiveness of the GAER model in community detection, rather than a trivial extension of the GAE model.

\begin{figure}[h]
  \centering
  \includegraphics[scale=0.5]{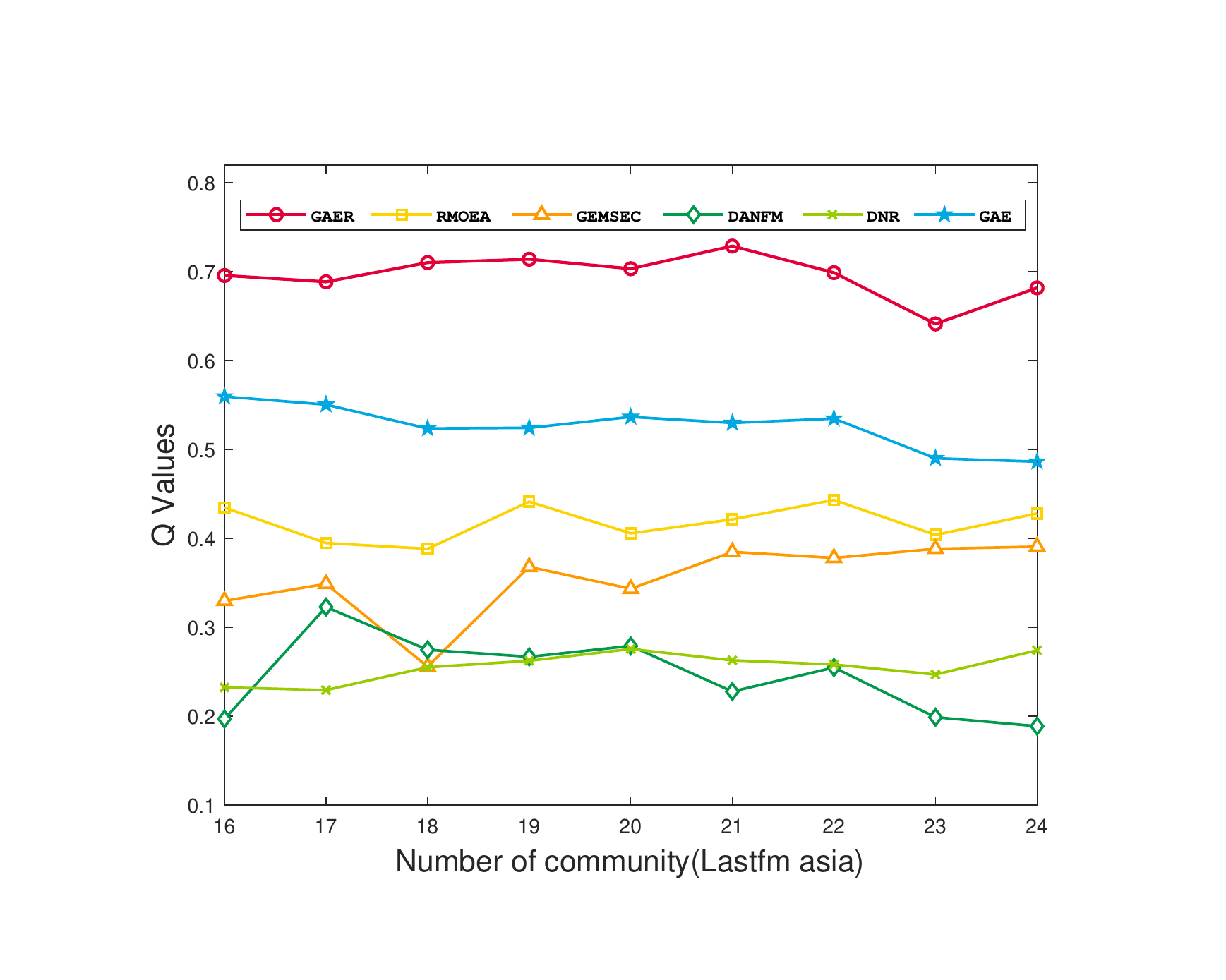}\\
  \caption{The $Q$ value of the algorithms on two networks with unknown community structure.}\label{ASIA}
\end{figure}

At last, we focused on analyzing a real social network $G_{12}$. It has always been an interesting topic to find close communities from social networks to guide marketing, relationship discovery, etc. At the same time, determining the number of communities in a real network generally has different results, so we set a optimal community range $G_{12}$, and experiments were carried out. The specific results are shown in the fig.\ref{ASIA}, GAER achieves excellent results near 0.7 under each number setting ($Q$ value on the real network is generally 0.3 to 0.7 caused by sparseness of the network edge), and other algorithms have average bad performance.
\begin{table*}[t]
\center
\caption{The time cost and accuracy at inference stage}
\label{Inference}
\begin{tabular}{|c|c|c|c|c|c|c|}
  \hline
  & & $\operatorname{GAER}_3$ & $\operatorname{GAER}_2$ & $\operatorname{GAER}_1$ & $\operatorname{GAER-APAM}_1$ & $\operatorname{GAER-APAM}_2$ \\
  \hline
  \multirow{2}{*}{Facebook}
    & \emph{Inference time} & 20.1534(1$\times$) & 1.3720(14.69$\times$) & 0.2436(82.73$\times$) & 0.4034(49.96$\times$) & 3.2762(6.15$\times$) \\
  \cline{2-7}
    & \emph{NMI} & \textbf{0.537} (-0.00\%)& 0.493 (-8.19\%) & 0.447 (-16.76\%) & 0.472 (-12.10\%) & \emph{0.516} (-3.91\%)\\
  \hline
  \multirow{2}{*}{AliGraph}
    &\emph{Inference time}  & 113.2765(1$\times$) & 3.8762(29.22$\times$) & 0.6954(162.89$\times$) & 0.9762(116.04$\times$) & 8.0756(14.03$\times$) \\
  \cline{2-7}
    & \emph{NMI} & \textbf{0.387} (-0.00\%)& 0.309 (-20.16\%) & 0.243 (-37.21\%)& 0.296 (-23.51\%)& \emph{0.369} (-4.65\%)\\
  \hline
\end{tabular}
\end{table*}

\subsection{Experiment on real-time large graph}
\begin{table}[h]
\center
\caption{Two real-time large graphs }
\label{Large}
\begin{tabular}{cccccc}
\hline\noalign{\smallskip}
Symbol & Dataset & K & N & M & C \\
\noalign{\smallskip}\hline\noalign{\smallskip}
$G_{13}$ & Facebook & 4 & 22470 & 171002 & K \\
$G_{14}$ & Aligraph & 7 & 46800 & 946175 & K \\
\noalign{\smallskip}\hline
\end{tabular}
\end{table}
In this part, we selected two representative large-scale networks Facebook and AliGraph for experiments. The network properties are shown in Table.\ref{Large}. We extract 20\% of the data as real-time data for inference detection, and the other 80\% as inherent offline data for community detection training.

\begin{figure}[h]
  \centering
  \includegraphics[scale=0.3]{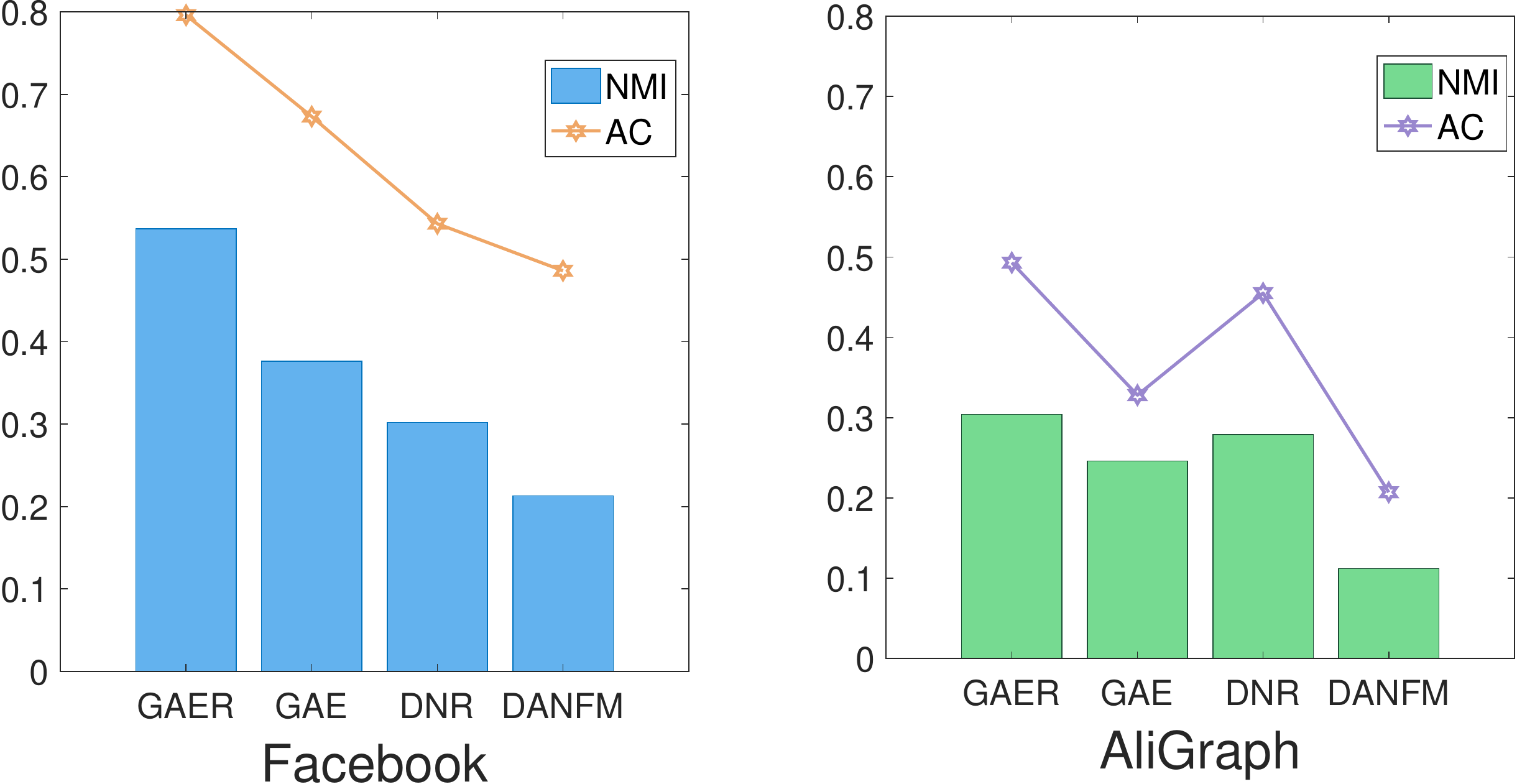}\\
  \caption{The NMI and AC results of four kinds of deep learning algorithms on two large-scale real networks.}\label{fig3}
\end{figure}

We also use GAER, GAE, DNR, and DANFM to conduct community detecting on 80\% of the offline data. GAER, GAE, DNR and DANFM are all three-layer structures, and the evaluation index is NMI and AC (accuracy). The results are shown in Fig.\ref{fig3}.

Then, after the training of each layer of GAER model is completed, we use 1-layer, 2-layer, and 3-layer GAER models and 1-layer and 2-layer GAER-APAM models for real-time data inference. And use the most time-consuming 3-layer GAER model as Benchmark, statistical time-consuming and acceleration multiple of each inference model. We also record the standardized mutual information NMI of each model for node data community detection, as shown in Table.\ref{Inference}. It should be noted that the bold NMI item in the table is ranked first, and the italicized item is ranked second.

As shown in the Table.\ref{Inference}, GAER-APAM compares GAER in the same layer to increase by 4.67\%-95.28\% in the community detection effect; and 2-layer GAER-APAM compares with the three-layer GAER, only a small accuracy (3.91\%-4.65\%) is sacrificed. But in exchange for a 6.15$\times$-14.03$\times$ time acceleration. To sum up, GAER-APAM is a method that takes into account both accuracy and time efficiency.

\section{Conclusion}
This paper proposes a novel community detection framework GAER based graph autoencoder reconstruction for the first time. Through modularity optimization and neighborhood Laplacian smooth sampling, by jointly using node modularity features and neighborhood high-quality information, it greatly improves the accuracy of GAER in community detection. At the same time, we combine the characteristics of the current big data system to design a two-stage GAER framework to further reduce the complexity of the algorithm and accelerate community detection. Finally we proposed an APAM module designed for the new nodes in the large-scale network, we can make faster inference while ensuring accuracy, which improves the scalability of GAER under the big data system. A series of experiments confirmed the accuracy, efficiency and speed of the community detection framework GAER we designed. In the future, we will further compress input features to achieve faster and efficient community detection.

\section*{Acknowledgment}
\small
The authors are grateful to the anonymous reviewers for their valuable suggestions for improving the manuscript. And this work was partially supported by National Natural Science Foundation of China under Grant 71871233, and Fundamental Research Funds for the Central Universities of China under Grant 2020XD-A01-2.

\end{document}